\documentclass[11pt,onecolumn, peerreview,draftclsnofoot]{IEEEtran}
 
\usepackage[cmex10]{amsmath}
\usepackage{url}
\usepackage{graphicx}
\usepackage{color}
\usepackage{placeins}
\usepackage{float}
\usepackage{tabularx,colortbl}
\usepackage{multirow}
\usepackage{xfrac}
\usepackage{array}
\usepackage [autostyle, english = american]{csquotes}
\usepackage[table]{xcolor}    

\begin{document}

\title{Extending LTE into the Unlicensed Spectrum: Technical Analysis of the Proposed Variants}

\IEEEoverridecommandlockouts

\author{\IEEEauthorblockN{Mina Labib*, Vuk Marojevic*, Jeffrey H. Reed*,  Amir I. Zaghloul*\textsuperscript{\textdagger} }
	\IEEEauthorblockA{\\ *Virginia Tech, Blacksburg, VA, USA\\
		\textsuperscript{\textdagger}US Army Research Laboratory, Adelphi, MD, USA}}

\maketitle

\begin{abstract}
The commercial success of the Long Term Evolution (LTE) and the resulting growth in mobile data demand have urged cellular network operators to strive for new innovations. LTE in unlicensed spectrum has been proposed to allow cellular network operators to offload some of their data traffic by accessing the unlicensed 5 GHz frequency band. Currently, there are three proposed variants for LTE operation in the unlicensed band, namely LTE-U, Licensed Spectrum Access (LAA), and MulteFire. This paper provides a comparative analysis of these variants and explains the current regulations of the 5 GHz band in different parts of the world. We present the technical details of the three proposed versions and analyze them in terms of their operational features and coexistence capabilities to provide an R\&D perspective for their deployment and coexistence with legacy systems.

\end{abstract}


\section{Introduction}

The demand in mobile traffic has been growing tremendously since the introduction of smartphones in 2007. Since then, cellular network operators have been looking for new technologies to meet the demand. At that time, the 4G Long Term Evolution (LTE) standard had almost been completed by the 3rd Generation Partnership Project (3GPP). LTE specifications were finalized by the 3GPP in March 2009 (3GPP Rel-8). Cellular network operators quickly deployed LTE, starting in December 2009 by the Swedish-Finnish operator (TeliaSonera), as the enabling technology to meet the demand for more wireless data. True to its namesake, LTE has been able to keep pace with the growing demand for capacity through several added features and modifications. 3GPP Rel-10 for LTE-Advanced (LTE-A) was finalized in June 2011 and fully meets the 4G system requirements. LTE-A includes features such as carrier aggregation (CA), which allows mobile operators to aggregate several frequency chunks into a larger bandwidth. In October 2015, the 3GPP announced the plan to further evolve LTE for paving the path towards 5G through LTE-Advanced Pro (LTE-A Pro). LTE-A Pro refers to LTE enhanced with the new features included in the 3GPP specifications starting from Rel-13 (which was finalized in March 2016), Rel-14 (which was released in January 2017 and expected to be finalized before September 2017), and onwards.

One of the salient features of LTE-A Pro is extending LTE into the 5 GHz unlicensed spectrum, comprising the frequency range between 5150 MHz and 5925 MHz. The 5 GHz band, which is also known as the U-NII (Unlicensed National Information Infrastructure) band, is currently utilized by various radar systems, in addition to Wireless Local Area Networks (WLAN), specifically the ones that are based on IEEE 801.11a/g/n/ac technologies, which are also referred to as Wi-Fi systems. 

Currently, there are three proposed variants of LTE in unlicensed bands \cite{qualcom2}. The first is called LTE-U and is developed by the LTE-U Forum to work with the existing 3GPP Releases 10/11/12. LTE-U was designed for quick launch in countries, such as the United States and China, that do not mandate implementing the listen-before-talk (LBT) technique. The second variant is Licensed Assisted Access (LAA) and has been standardized by the 3GPP in Rel-13. LAA adheres to the requirements of the LBT protocol, which is mandated in Europe and Japan. It promises to provide a unified global framework that complies with the regulatory requirements in the different regions of the world. Both variants, LTE-U and LAA, use licensed spectrum as the primary carrier for signaling (control channels) and to deliver data for users with high Quality-of-Service (QoS) requirements. Carrier aggregation is used to add secondary component carriers in the unlicensed spectrum to deliver data to users with best-effort QoS requirements. MulteFire is the third variant of LTE in unlicensed bands and has been proposed as a standalone version of LTE for small cells. This variant will use only the unlicensed spectrum as the primary and only carrier, and it will provide an opportunity for neutral hosts to deploy LTE in the future. 

As opposed to other survey papers on this topic, such as \cite{lteu_survey} and \cite{lteu_wifi_survey}, this paper identifies the motivation for introducing different modes of LTE operation in the unlicensed spectrum, analyzes them with respect to regulations and coexistence capabilities, and identifies research issues and the way forward. The rest of the paper is organized as follows: Section II summarizes the current regulations for the unlicensed 5GHz band in the different parts of the world as the basis for our analysis. Section III provides a general overview of extending LTE into the unlicensed spectrum in terms of potentials and challenges. Sections IV-VI discuss the technical details of each of the proposed variants of LTE for unlicensed band operation. We present numerical results in Section VII, comparing the performance of the three variants in terms of coexistence with Wi-Fi. Section VIII provides the conclusions and an R\&D perspective on future deployment and coexistence of radios in unlicensed spectrum.

\section{Regulations in the 5 GHz Band}
The regulatory requirements to access the spectrum are different in every region in the world, and the same applies for the 5 GHz band. In general, there are several compliance rules that have been defined around the world for regulating the use of the unlicensed spectrum. These rules can be summarized as follows:

\begin{itemize}
	\item Limitations of the maximum transmission power and the maximum power spectrum density (PSD). 
	\item Use indoor only or use both indoor and outdoor. 
	\item Dynamic Frequency Selection (DFS).
	\item Listen-Before-Talk (LBT).
	\item Transmission Power Control (TPC).
\end{itemize}

DFS is a mechanism that is specifically designed to avoid causing interference to non-IMT (International Mobile Telecommunications) systems, such as radars. According to the Federal Communications Commission (FCC) regulations in the United States, any device (working in certain sub-bands of the 5 GHz band) must sense the channel before using it, and sense it periodically to ensure there is no radar system using this channel. If a radar signal is detected, i.e. the received power levels is above a certain threshold, the operating channel must be vacated. The device must not utilize that channel for the non-occupancy period of thirty minutes \cite{fcc1}. 

LBT is a mechanism introduced for fair co-existence with other wireless communication systems (such as Wi-Fi). In Europe and Japan, there is a mandatory requirement to implement LBT when accessing the unlicensed spectrum. According to the European Telecommunications Standards Institute (ETSI), and based on the Load Based Equipment rules, any device that wants to access the unlicensed spectrum, needs to perform Clear Channel Assessment (CCA) beforehand, which translates to spectrum sensing for certain period (called CCA period and it is greater than 20 $\mu$s). If  the detected energy is lower than a certain threshold (which equals -73 dBm/MHz for the case of a transmitter with an EIRP of 23 dBm and assuming receiver antenna gain of 0), the device can access the channel for a period called channel occupancy time (which should be less than \(\frac{13}{32}\)q, where q is selected by the manufacturer and it is in the range 4-32). Then the device has to stay idle for a minimum period of CCA multiplied by a number that is randomly selected between 1 and q \cite{3gpp.36.889}. 

Table \ref{rules} captures the regulatory requirements in the major regions of the world \cite{3gpp.36.889}. The maximum transmission power in every sub-band is limited by regulatory requirements, which motivates designing the LTE in unlicensed bands for the small cell network deployment. 

\begin{table}[]
	\fontsize{8}{8}\selectfont
	\centering
	\caption{Regulatory requirements in Different World Regions.}
	\label{rules}

	\begin{tabular} {|p{0.9cm}|m{0.6cm}|m{2cm}|m{2cm}|m{1.2cm}|m{2cm}|m{2cm}|m{2cm}|}
		%
	
	\hline
	&       & \textbf{5150-5250 MHz}    & \textbf{5250-5350 MHz} & \textbf{5350-5470 MHz} & \textbf{5470-5725 MHz} & \textbf{5725-5850 MHz} & \textbf{5850-5925 MHz}   \\ 
	\hline
	 & Usage & WAS/RLAN   & WAS/RLAN  & Under consideration  &  WAS/RLAN & FWA (Allowing WAS is under consideration)   & ITS (Allowing WAS is under consideration) \textsuperscript{(1)} \\ 	
	\cline{2-4} \cline{6-8}
    \multirow{-2}{0.7cm}{\textbf{Europe}}& Rules & Indoor only,  max Tx Power is 23 dBm, max PSD is 10 dBm/MHz, No TPC,  No DFS, LBT  & Indoor only, TPC (mx EIRP is 23 dBm, max PSD is 10 dBm/MHz), DFS, LBT &    & Indoor/Outdoor, TPC (max EIRP is 30 dBm, max PSD is 17 dBm/MHz), DFS, LBT  & Indoor/Outdoor, TPC (max EIRP is 33 dBm, max PSD is 23 dBm/MHz), DFS, No LBT  & Indoor/Outdoor, TPC (max EIRP is 36 dBm),  DFS (none for 5850-5875 MHz), LBT \\
	 \hline
	  & Usage & U-NII-1  & U-NII-2A    & U-NII-2B  & U-NII-2C  & U-NII-3    & U-NII-4   \\ 
	\cline{2-4} \cline{6-7}
	 \multirow{-2}{*}{ \textbf{USA} } & Rules & Indoor/Outdoor, max Tx Power is 30 dBm, max PSD is 17 dBm/MHz, No TPC,  No DFS, No LBT & Indoor/Outdoor, max Tx Power is 24 dBm, max PSD is 11 dBm/MHz, TPC, DFS, No LBT & (Under consideration) & Indoor/Outdoor, max Tx Power is 24 dBm, max PSD is 11 dBm/Mhz, TPC, DFS, No LBT & Indoor/Outdoor, max Tx Power is 30 dBm, max PSD is 30 dBm in 500 KHz, No TPC, No DFS, No LBT & (Under consideration) \\ 
	\hline
     & Usage & RLAN     & RLAN  &   & RLAN \textsuperscript{(2)}  & RLAN  &  \\
	 \cline{2-4} \cline{6-7}
     \multirow{-2}{0.7cm}{\textbf{Canada}}& Rules & Indoor only, max EIRP 200 mW, max PSD is 4 dBm/MHz, No TPC,  No DFS, No LBT  & Indoor/Outdoor, max Tx Power is 24 dBm, max PSD is 11 dBm/MHz, TPC, DFS, No LBT & N/A   & Indoor/Outdoor, max Tx Power is 24 dBm, max PSD is 11 dBm/MHz, TPC, DFS, No LBT   & Indoor/Outdoor, max Tx Power is 30 dBm, max PSD is 17 dBm/MHz, No TPC, No DFS, No LBT & N/A \\
	 \hline
	 & Usage & RLAN  & RLAN &  & RLAN  & RLAN    & \\ 
	\cline{2-4} \cline{6-7}
	 \multirow{-2}{0.7cm}{\textbf{Brazil} }& Rules & Indoor only, max EIRP 200 mW, max EIRP PSD is 10 mW/MHz, no DFS  & Indoor only, max EIRP 200 mW,  max EIRP PSD is 10 mW/MHz, DFS  & N/A   & Indoor/Outdoor, max Tx power is 250 mW,  max EIRP PSD is 50 mW/MHz, DFS  & Indoor/Outdoor, max Tx power is 1 W   &  N/A \\
	\hline
	 & Usage & WAS/RLANs   & WAS/RLANs  &  &   &  RLAN  & \\ 
	\cline{2-4} \cline{7-7}
	 \multirow{-2}{0.7cm}{\textbf{China} } & Rules & Indoor only, max EIRP 200 mW, max EIRP PSD is 10 dBm/MHz, TPC, DFS, no LBT   & Indoor only, max EIRP 200 mW, max EIRP PSD is 10 dBm/MHz, TPC, DFS, no LBT & N/A &  Under Consideration & Rules not formally issued  & N/A	\\ \hline
	 & Usage & RLAN    & RLAN      &   & RLAN   &   & \\ 
	\cline{2-4} \cline{6-6}
	 \multirow{-2}{0.7cm}{\textbf{Japan}}  & Rules & Indoor only, max Tx power depends on BW, no TPC, no DFS, LBT    & Indoor only, max Tx power depends on BW, TPC, DFS, LBT  & N/A & Indoo/Outdoor, max Tx power depends on BW, TPC, DFS, LBT     &  N/A &  N/A  \\ 
	\hline
	\multicolumn{8}{|l|}{(1) In Europe, FWA is utilizing the 5725-5875 MHz range, and ITS is utilizing the 5855-5925 MHz range.}\\
	\multicolumn{8}{|l|}{ (2) In Canada, RLAN is forbidden in the frequency range 5600-5650 MHz.} \\
	\multicolumn{8}{|l|}{\textbf{Abbreviations:} \hspace{0.3cm}  \textbf{WAS}: Wireless Access Systems, \textbf{RLAN}: Radio Local Area Networks, \textbf{FWA}: Fixed Wireless Access,}\\
	\multicolumn{8}{|l|}{  \hspace{2.2cm}  \textbf{U-NII}: Unlicensed National Information Infrastructure. }\\

	\hline


\end{tabular}

\end{table}

\section{Benefits and Challenges of LTE in Unlicensed Bands}

Extending the use of LTE in the 5-GHz unlicensed band can achieve several benefits compared to Wi-Fi. The benefits can be summarized as \cite{lteu_survey}:

\begin{itemize}
	\item Better spectrum efficiency: LTE is using scheduled-based channel access compared to the contention-based scheme used by Wi-Fi. That leads to offering more efficient multiuser channel access and improves improve system capacity. Recent simulation results have shown that both LTE-U and LAA can achieve twice the capacity offered by Wi-Fi \cite{qualcom2}. 
	\item Larger coverage area: LTE uses the more effective 1/3 turbo coding to overcome low SINR, has a more robust control channel design, and implements the HARQ (Hybrid Automatic Repeat Request) protocol, which makes it more robust to interference.
	\item Unified LTE network: Operators will be able to use a single platform for authentication, registration and management. 
	\item Security: LTE offers more security than Wi-Fi due to the enhanced authentication procedures. 
	\item Better user experience: LTE offers good  mobility management (which is not well supported in Wi-Fi), so user will experience less service interruptions during mobility. Furthermore, the switch from using LTE in the licensed bands to the unlicensed one will be transparent to the users.  
\end{itemize}

However, LTE faces several challenges if it is extended into the 5 GHz unlicensed band in terms of coexistence with both radar and Wi-Fi systems. Wi-Fi systems are widely deployed in the 5 GHz band, and it is crucial that LTE in unlicensed bands does not cause degradation of Wi-Fi performance. Wi-Fi  was designed to operate in the unlicensed spectrum and employs Carrier Sense Multiple Access with Collision Avoidance (CSMA/CA) to access the channel and to ensure fair coexistence with other technologies. 

Recently, the coexistence between LTE and Wi-Fi in unlicensed bands has become an important area of research, from both industry and academia. The authors in \cite{perfEvalErica} have conducted system-level analysis to evaluate the performance of both LTE and Wi-Fi when working in the same band. They concluded that, if no modifications were done to LTE, the performance of Wi-Fi may be severely degraded, while the performance of LTE would remain almost unchanged. The authors in \cite{Almeida2013_ABS} suggest using the almost-blank subframe (ABS) feature, which was introduced in LTE Rel-10, to improve the coexistence between LTE and Wi-Fi technologies. ABS is a feature that allows the LTE base station (eNodeB) to transmit subframes that contain only the basic system information messages, and it is used for better coordination between macro-cells and small-cells. It can be considered as a sort of static muting mechanism. The authors in \cite{jeon2014a} show that a simple LBT algorithm will provide better coexistence performance than the static muting algorithm. Reference \cite{lteu_wifi_survey} surveys the coexistence mechanisms proposed for LTE and Wi-Fi systems in unlicensed bands. 

\section{LTE-U}
LTE-U is developed by the LTE-U Forum, which was formed in 2014 by Verizon, Alcatel-Lucent, Ericsson, Qualcomm Technologies Inc., and Samsung Electronics. The goal of LTE-U is to use the existing features in the latest LTE 3GPP specifications and adapt them to unlicensed operation in countries, such as the US and China, that do not mandate LBT. LTE-U supports supplemental downlink (SDL) only within the frequency bands 5150-5250 MHz and 5725-5850 MHZ, whereas the frequency bands 5250-5725 MHz has been reserved for future use. The last set of specifications for LTE-U (issued in October 2015) provides general technical guidelines and benchmarks for testing scenarios, but does not specify certain implementation mechanisms \cite{ltu_forum13}. 

LTE-U specifications were designed for the case when a single eNodeB has access to licensed spectrum (called Primary Cell or PC) and unlicensed spectrum (called Secondary Cell or SC). Without modifying the 3GPP LTE specifications, there are several mechanisms that are used for LTE-U to better coexist with Wi-Fi \cite{qualcom2}, such as carrier selection, on-off switching, and Carrier-Sensing Adaptive Transmission (CSAT).

\subsection{Carrier Selection}
The eNodeB performs carrier selection at startup, periodically and based on performance triggers. Carrier selection implies scanning the spectrum and measuring the power level in each channel to find the channel that is free of interference. If all channels are occupied by other systems, the eNodeB choses the channel with the lowest detected signal power level. The eNodeB will continue monitoring channel activities and select a more suitable channel when available. Carrier selection algorithm was left to be implementation specific. 

\subsection{On-Off Switching}
When the traffic demand is low, the small cell eNodeB can stop transmitting in the unlicensed spectrum and relies only on the licensed spectrum. Doing this will reduce the amount of interference to Wi-Fi users. LTE-U specifications define two states for the LTE-U SC \cite{ltu_forum_csat}:
\begin{itemize}
	\item Off-State: The SC stops any type of transmission.
	\item On-States: The SC is either transmitting full LTE frames according to the 3GPP specifications or transmitting the LTE-U Discovery Signal (LDS). The LDS is transmitted by the SC at a certain subframe (subframe number 5) and with fixed time intervals defined by the LDS periodicity parameter (which can be either 40, 80, or 160 ms). LDS contains the physical signals and channels required for the LTE User Equipment (UE) to obtain time and frequency synchronization and to perform SC measurements.

\end{itemize} 

\subsection{ Carrier-Sensing Adaptive Transmission}
 CSAT is a mechanism that allows the eNodeB to share the spectrum with other systems using the same channel in a Time Division Multiplex (TDM) manner. Qualcomm proposed CSAT as a spectrum sharing technique to be used with LTE-U \cite{qualcom2}. CSAT, in concept, is a sort of adaptive muting algorithm, where the eNodeB initially and periodically senses the channel for relatively long time periods (anywhere in the rage between 0.5 and 200 ms). As a function of the channel activity and detection of Wi-Fi signals above the energy threshold level (which is -62 dBm), SC will adjust its duty cycle and define a time cycle for transmission. Since Wi-Fi stations use carrier sensing, they will be able to adjust their own transmissions in the periods when the duty cycle of LTE-U SC is off. The CSAT duty cycle can change over time based on channel usage, but the constraint values are \cite{ltu_forum_csat}:
  
\begin{itemize}
  	\item Minimum Off-State Period: 1 ms.
  	\item Maximum Off-State Period: Determined by the LDCS periodicity, which can be either 40, 80, or 160 ms. 
  	\item Minimum On-State Period: 4 ms in case of available user data and 1 ms (LDS period), otherwise..
  	\item Maximum On-State Period: 20 ms.
  	\item Energy detection threshold: -62 dBm. 
  	
\end{itemize}

An obvious drawback of CSAT is the long latency that may not be suitable for real-time applications over Wi-Fi. Another drawback is that the eNodeB can extend its transmission until the duty cycle reaches 90\%  when it cannot detect the Wi-Fi signal (hidden node problem), which will lead to diminishing the Wi-Fi signal.
  
Recently, the FCC approved that Qualcomm and Verizon perform small-scale testing for LTE-U in real-world scenarios at two different locations. If test results show that LTE-U can coexist fairly with Wi-Fi systems, LTE-U might be commercially deployed in 2017.

\section{Licensed Assisted Access (LAA)}

Licensed Assisted Access (LAA) was standarized in 3GPP Rel-13. The operating frequency band for LAA spans the frequency range 5150 MHz - 5925 MHz (channel numbers 46 and 47 in the 3GPP specifications). The current allowable bandwidths for LAA operation in unlicensed spectrum are 10 and 20 MHz \cite{3gpp.36.101}. Rel-13 defines LAA only for the downlink (DL). One of the main features of 3GPP Rel-14 is the introduction of enhanced-Licensed Assisted Access (eLAA), which includes uplink (UL) operation for LAA. 
LAA has a different frame structure type (Frame Sturcture Type 3) for operation in the unlicensed band. Frame Structure Type 3, similar to the one defined for FDD, has a duration of 10 ms and consists of 20 slots, each slot duration is 0.5 ms and each two adjacent slots form one subframe. Any subframe may be available for DL or UL transmission, however the transmission may or may not start at the boundary of the subframe, and it may or may not end at the boundary of the subframe \cite{3gpp.36.211}. This is due to the fact that the eNodeB has to sense the spectrum before transmitting and transmits only if the channel is free.  Since one of the goals of LAA is ensuring fair coexistence with Wi-Fi, we will further examine the channel access procedures for eNodeB in the unlicensed spectrum as detailed in \cite{3gpp.36.213}.

3GPP Rel-13 has identified two different modes for the eNodeB to transmit in the unlicensed spectrum. The first mode is for transmitting the Physical Downlink Shared Channel (PDSCH), which is the channel that carries user data on the DL. The second mode is for transmitting a Discovery Reference Signal (DRS) without the PDSCH. 

\begin{figure}[t]
	\centering
	\begin{center}
		\includegraphics[width=5in]{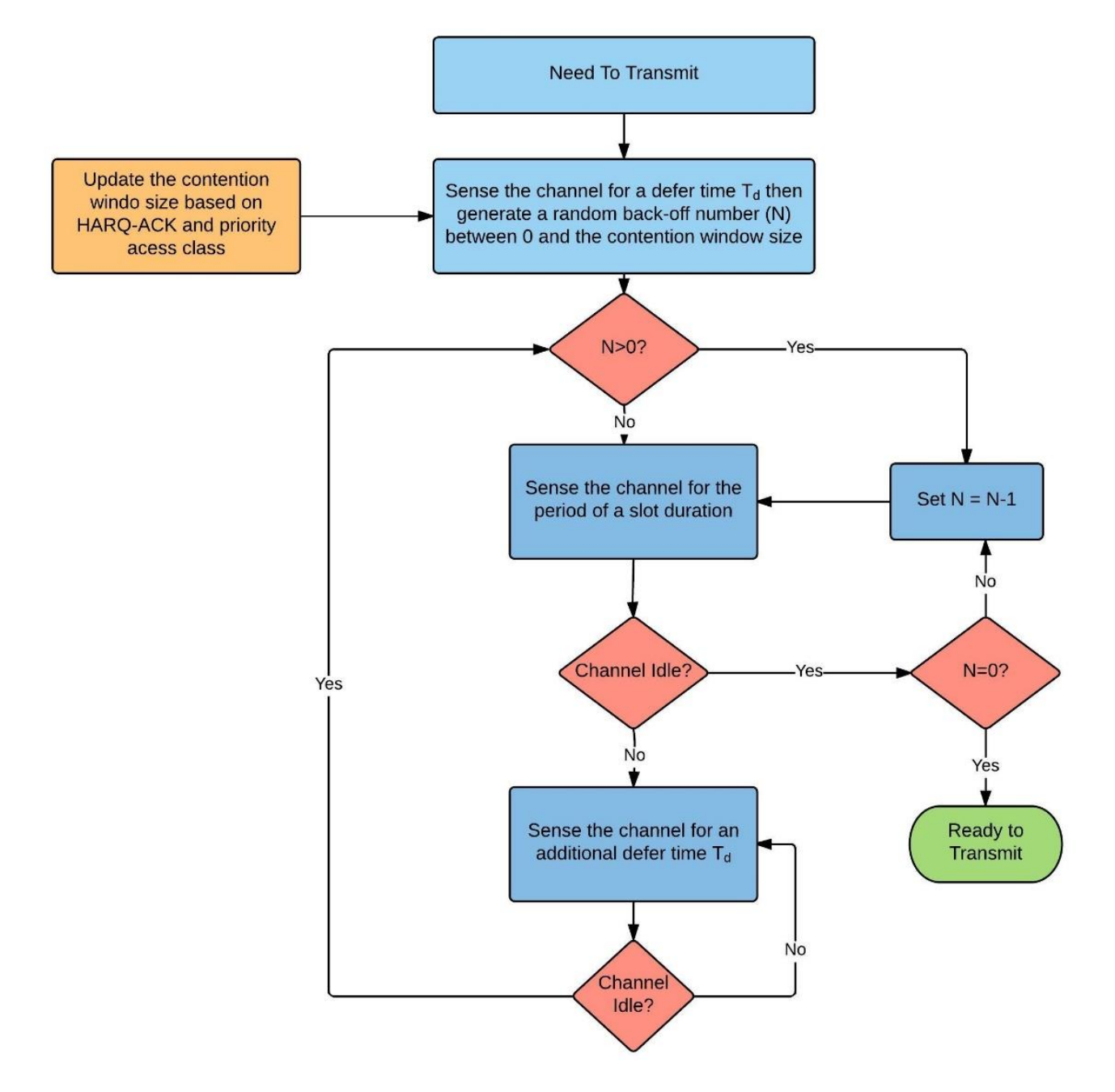} 
		\caption{eNodeB procedures to access the unlicensed spectrum before transmitting PDSCH.}
		\label{laa_access}
	\end{center}
	\vspace{-0.25in}
\end{figure}

For the case of transmitting the PDSCH, Figure \ref{laa_access} illustrates the main procedures that need to be followed by the eNodeB. These procedures are based on the LBT algorithm with random back-off and a variable contention window size (LBT Category 4), where the eNodeB generates a random number within the contention window size to identify the period it needs to sense the spectrum before transmission. The size of the contention window is variable as the device will increase the window size when it finds the spectrum occupied. The 3GPP specifications have identified four different priority access classes, which define the channel sensing parameters. It is worth mentioning that these parameters are very similar to those used for the different access categories in IEEE 802.11, with the aim of achieving fair coexistence among the two technologies.

To elaborate more, the eNodeB starts by sensing the channel for a period called defer duration time \(T_d\), then selects a random number \(N\) that is uniformly distributed between 0 and \(CW_p\), where \(CW_p\) is the contention window size that has a minimum and maximum values, which  are determined by the channel access priority class. The eNodeB senses the channel for an additional period of \(N+1\) times the slot duration, which equals 9 $\mu$s. If the channel is found to be idle during all the slot durations of \(T_d\) and during the \(N+1\) slot durations, the eNodeB can start transmitting with a maximum duration of \(T_{mcot,p}\) (which ranges between 2 and 10 ms depending on the priority channel access class). The channel is determined to be idle during a certain time slot if the detected power is less than a certain energy detection threshold \(X_{Thresh}\) for at least 4 $\mu$s of the slot duration time. If the channel is found to be busy during any time slot, the eNodeB continues sensing the channel for an additional defer duration time \(T_d\). The defer duration time \(T_d\) consists of a duration \(T_f\), which equals 16  $\mu$s, followed immediately by \(m_p\) additional slot durations \(T_{sl}\). Each slot duration  \(T_{sl}\) is 9 $\mu$s and \(T_f\) includes an idle slot duration \(T_{sl}\) at its start, wile \(m_p\) ranges between 1 and 7, where its value depends on the priority channel access class.


For the choice of the contention window size \(CW_p\), the eNodeB starts with the minimum value \(CW_{min}\) and if the data was not received correctly by the users, the eNodeB chooses the next higher \(CW_p\), and so on until reaching the maximum allowable value \(CW_{max}\) and continues using it until the data is correctly received.

For the case of transmitting a DRS, the eNodeB needs to sense the channel for a period of 25 $\mu$s and if the channel is found to be idle for the entire period, the eNodeB can transmit a discovery signal for a maximum period of 1 ms. The DRS is identified for Frame Structure Type 3 and contains 12 OFDM symbols within one non-empty subframe. It carries the primary synchronization signal (PSS), secondary synchronization signal (SSS) and cell specific reference signal (CRS) \cite{3gpp.36.211}.

The value of the energy detection threshold \(X_{Thresh}\) is determined differently in two different cases. For the case where there is no other technology occupying the channel on a long term basis (as in the case of regulations in certain regions) and for 20 MHz bandwidth,

\begin{equation} \label{eq:fc1}
X_{Thresh} = min\begin{cases}
-52 dBm,    \\ X_r   \\
\end{cases}
\end{equation}
where \(X_r\) is the maximum energy detection threshold defined by regulation requirements in dBm (if defined).

For the case where multiple technologies are allowed to share the channel and for the 20 MHz bandwidth, 

\begin{equation} \label{eq:fc2}
X_{Thresh} = max\begin{cases}
-72 dBm,    \\ min \begin{cases} 
-62 dBm, \\ -62 - T_A + (23- P_{Tx}) dBm, \\   \\
\end{cases}
\end{cases}
\end{equation} \\
where \(T_A\) is either 10 dBm in case of transmitting PDSCH or 5 dBm in case of transmitting DRS, and \(P_{T_x}\) is the maximum eNodeB output power in dBm. 

For UL transmission, which was defined in Rel-14, the UE also is allowed to use one of two modes to transmit in the unlicensed spectrum. These two modes of sensing the spectrum are similar to the ones used by the eNodeB. In most cases, the eNodeB informs the UE of the mode that it needs to use before accessing the spectrum along with the UL grant.  
It is worth mentioning that the 3GPP specifications have also identified the procedures for the eNodeB when accessing multiple channels in the unlicensed spectrum, allowing the eNodeB to choose the channel that has that lowest detected signal power level to reduce interfering with other existing systems within the unlicensed spectrum. 
Qualcomm has performed a laboratory tests to evaluate the performance of the coexistence algorithm implemented in LAA. The results show that when an operator switches from Wi-Fi to LAA, the throughput for this operator increases by 100\%, and even the throughput for the Wi-Fi operators increases by approximately 10\% \cite{qualcom2}.

\section{MulteFire}
MulteFire was proposed as a standalone version of LTE for small cells. In December 2015, MulteFire alliance was formed by Qualcomm, Nokia, Ericsson and Intel to promote the MulteFire technology and several companies have joined the alliance since then. The first release of the technical specifications for MulteFire was issued in Jan. 2017 \cite{multefire1}. MulteFire, similar to Wi-Fi, relies only the unlicensed spectrum and can provide service to users with or without USIM (Universal Subscriber Identity Module) card. Hence, MulteFire will combine the benefits of the advanced LTE technology and the simplicity of Wi-Fi deployment \cite{qualcom2}. 

MulteFire can be deployed either by traditional mobile operators or by neutral hosts. Accordingly, MulteFire specifies two different architectures: 

\begin{itemize}
	\item Public Land Mobile Network (PLMN) access mode, which allows mobile network operators to extend their coverage into the unlicensed band, specially in case where licensed spectrum is not available at certain locations.  
	\item Neutral Host Network (NHN) access mode, which is similar to Wi-Fi, a self-contained network deployment that provides access to the Internet. 
\end{itemize}

Because of the nature of transmission in the unlicensed band and the need to adhere to the LBT requirements, MulteFire has introduced several modifications in the radio air interface compared to  LTE.

\subsection{Downlink Operation}
As for LAA, a MulteFire eNodeB will need to perform LBT before transmitting any signal. The LBT procedure is similar to the one of LAA and has the same four channel access priority classes. Also similar to LAA, the eNodeB can transmit a DRS that contains critical data for synchronization and acquiring the system information. DRS for MulteFire is also 12 OFDM symbols long, but its structure is different when compared to LAA. The components of DRS for MulteFire are:

 \begin{itemize}
 	\item Primary Synchronization Signal (PSS): transmitted on the seventh OFDM symbol of the DRS subframe. 
 	\item Secondary Synchronization Signal (SSS): transmitted on the sixth OFDM symbol of the DRS subframe. 
 	\item MulteFire Primary Synchronization Signal (MF-PSS): transmitted on the fourth OFDM symbol of the DRS subframe. 
 	\item MulteFire Secondary Synchronization Signal (MF-SSS): transmitted on the third OFDM symbol of the DRS subframe. MF-PSS and MF-SSS support the UE in performing frequency/time synchronization and also allow differentiating between an LAA eNodeB and a MulteFire eNodeB. 
 	\item Cell-specific reference signals (CRS). 
 	\item Configurable channel state information reference signals (CSI-RS).
 	\item Master information broadcast (MIB-MulteFire) through the MulteFire Physical Broadcast Channel (MulteFire-PBCH), which is transmitted over six OFDM symbols (the fifth, eighth, ninth, tenth, eleventh and last OFDM symbol). 
 	\item MulteFire system information broadcast (SIB-MulteFire) which is transmitted through PDSCH and carries information similar to the SIB1 and SIB2 messages of LTE Rel-13. 
 \end{itemize}

 MulteFire allows sending the DRS in two modes:
  
   \begin{itemize}
   	\item During the serving cell DRS measurement and timing configuration (DMTC) window, which can be up to 10 ms long and during which the UE expects to receive the DRS. The DMTC periodicity is 40, 80 or 160 ms. MulteFire will transmit the DRS during the DMTC window after sensing the channel for a period of 25 $\mu$s.
   	
   	\item Just like for the downlink PDSCH, opportunistic transmission of DRS is allowed only on subframe number 0, after performing the LBT mechanism.

   \end{itemize}

\subsection{Dynamic DL/UL Configuration}
MulteFire adopts a very flexible frame structure to dynamically adapt to the DL and UL traffic loads. Accordingly, the ratio between DL and UL transmission can vary from one frame to the next.  The eNodeB will broadcast whether a subframe is DL or UL through the Common Physical Downlink Control Channel (C-PDCCH).

\subsection{Uplink Operation}
MulteFire uses Block Interleaved FDMA (B-IFDMA) as the UL transmission scheme, where the bandwidth is divided into N interlaces (N = 10 for 20 MHz, and N = 5 for 10 MHz), each interlace consists of 10 equally spaced physical resource blocks. 
MulteFire introduces two different formats for the Physical Uplink Control Channel (PUCCH): Short-PUCCH (MF-sPUCCH) and Extended PUCCH (MF-ePUCCH). MF-sPUCCH is transmitted by the UE during the last four OFDM symbols of a DL subframe. In other words, the UE is allowed to transmit the MF-sPUCCH immediately on the gap between DL and UL transmission, and, therefore, does not need to perform LBT. This is allowed according to the ETSI regulations in the unlicensed band, as long as the UE is transmitting within 16 $\mu$s after the DL transmission. Due to its compact design, MF-sPUCCH will be used to carry small control information such as reception acknowledgments. On the other hand, MF-ePUCCH will be used by the UE to transmit large control information such as the channel state information (CSI). The UE will transmit MF-ePUCCH based on the UL resource assignment given by the eNodeB, the same way the UE transmits the Physical Uplink Shared Channel (PUSCH). 

Because of the huge demand for high data rates and for higher capacity, MulteFire has the potential to use spectrum opportunities and leverage LTE-A techniques to access the unlicensed spectrum. Furthermore, MulteFire supports mobility management and can provide a better user experience than Wi-Fi. 
However, MulteFire faces several challenges. Wi-Fi is already very widely used nowadays, not only in cellphones, but also in portable computing devices (laptops), so Wi-Fi chips are manufactured at low cost because of it high penetration. Furthermore, MulteFire will not be backward compatible  with legacy LTE devices. Moreover, there are still lots of design issues need to be considered, such as the control channel design and their reliability in the unlicensed band. One of these issues is handling the unintentional interference generated among several eNodeBs in dense deployment as illustrated in \cite{Mina_Mag1}.



\section{Comparative Analysis}

\begin{figure}[h]
	\centering
	\begin{center}
		\includegraphics[width=7in]{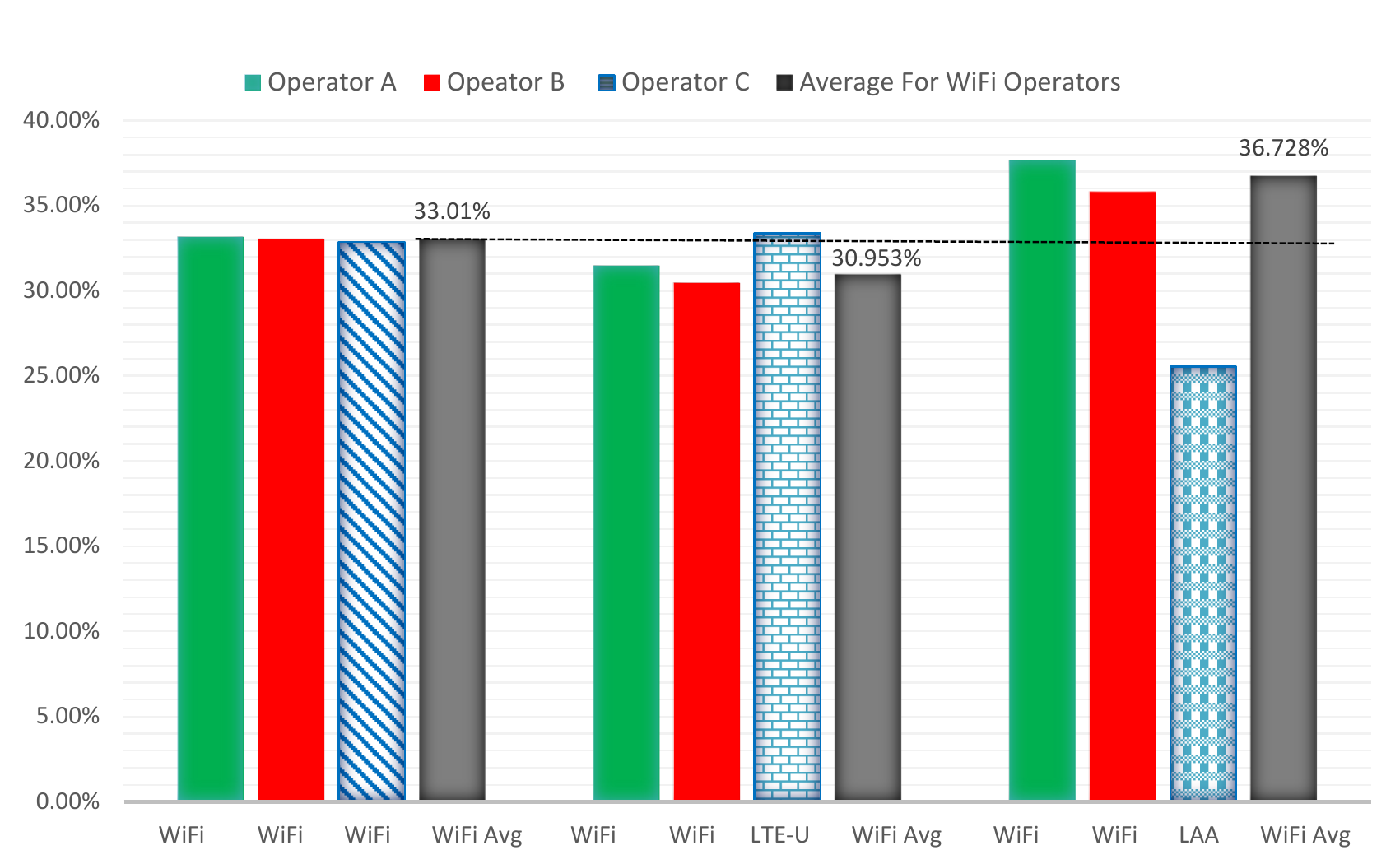} 
		\caption{Performance comparison in terms of normalized total time of transmission opportunities (free of collision).}
		\label{fig:techcompare}
	\end{center}
	\vspace{-0.25in}
\end{figure}

The objective of this analysis is to compare the coexistence performance of CSMA/CA (used in Wi-Fi), LBT (used in LAA and MulteFire), and CSAT (used in LTE-U). The performance is compared based on the normalized total time of transmission opportunities (TTTO) that is  granted (free of collision) to an operator, who is either deploying Wi-Fi, LAA/MulteFire, or LTE-U. Table \ref{simulation_parameters} shows the simulation parameters. TTTO is calculated based on the summation of transmission opportunity durations that an operator gets free of collision, normalized by the total analysis time. The coexistence performance of LAA/MulteFire or LTE-U is evaluated based on the average TTTO that a Wi-Fi operator will get (when adding an LAA/MulteFire or LTE-U eNodeB) compared to the case when adding another Wi-Fi operator. The simulation results are presented in Figure \ref{fig:techcompare} and illustrate the following: When deploying three Wi-Fi operators, the average TTTO per operator is 33.01\%. When one operator switches to LAA, the TTTO for each of the remaining operators increases to 36.728\% on average. When one of three Wi-Fi operators switches to LTE-U, this TTTO decreases to 30.95\%. This means that LAA and MulteFire are better neighbors to Wi-Fi than Wi-Fi itself and than LTE-U. It is worth mentioning that the TTTO metric can be a good indicator of DL throughput for operators with the same technology, but this is not true for operators using different technologies. The throughput in this case will depend on the spectral efficiency of each technology and how efficient each technology can utilize the time when granted access to the channel. Although LAA and MulteFire perform the same in terms of TTTO, the throughput will be different for the two variants, since LAA uses the licensed spectrum to transmit the control messages.

In our simulations, it was assumed that the received power of the other operators is always above the detection threshold level. In reality that may not always be true. Actually, since the detection threshold level is -62 dBm for LTE-U and -72 dBm for LAA/MulteFire, LAA/MulteFire operators will have larger detection range for Wi-Fi signal and outperform LTE-U even more in terms of fair coexistence.

\begin{table}[]
	\centering
	\caption{Simulation Parameters}
	\label{simulation_parameters}
\begin{tabular}{|c|c|c|}
	\hline
	\multirow{3}{*}{General Simulation Parameters}                   & \multicolumn{1}{c|}{Sampling Time}                  & \multicolumn{1}{c|}{1 $\mu$s}    \\ \cline{2-3} 
	& \multicolumn{1}{c|}{Simulation Time}                & \multicolumn{1}{c|}{100 s}       \\ \cline{2-3} 
	& \multicolumn{1}{c|}{Traffic Model}                  & \multicolumn{1}{c|}{Full Buffer} \\ \hline \hline
	\multirow{8}{*}{Wi-Fi Specific Parameters}                       & Access Category                                     & Video                            \\ \cline{2-3} 
	& Minimum Contention Window                           & 7                                \\ \cline{2-3} 
	& Maximum Contention Window                           & 15                               \\ \cline{2-3} 
	& Arbitration Inter-Frame Spacing Number              & 2                                \\ \cline{2-3} 
	& Short Inter-Frame Spacing (SIFS)                          & 16 $\mu$s                        \\ \cline{2-3} 
	& DCF Inter-Frame Spacing (DIFS)                            & 34 $\mu$s                        \\ \cline{2-3} 
	& Data transmission time (based on maximum PPDU frame) & 5.484 ms                         \\ \cline{2-3} 
	& Acknowledgment Transmission Time                   & 34 $\mu$s                        \\ \hline \hline
	\multirow{5}{*}{LAA Specific Parameters}                         & Priority Class                                      & 2                                \\ \cline{2-3} 
	& Minimum Contention Window                           & 7                                \\ \cline{2-3} 
	& Maximum Contention Window                           & 15                               \\ \cline{2-3} 
	& \(m_p\)                                               & 1                                \\ \cline{2-3} 
	& \(T_{mcot,p}\)                                       & 3 ms                             \\ \hline \hline
	\multicolumn{1}{|c|}{\multirow{2}{*}{LTE-U Specific Parameters}} & On-State Period                                     & 12 ms                            \\ \cline{2-3} 
	\multicolumn{1}{|c|}{}                                           & Off-State Period                                    & 24 ms                            \\ \hline 
\end{tabular}
\end{table}

\begin{figure}[h]
	\centering
	\begin{center}
		\includegraphics[width=3.7in]{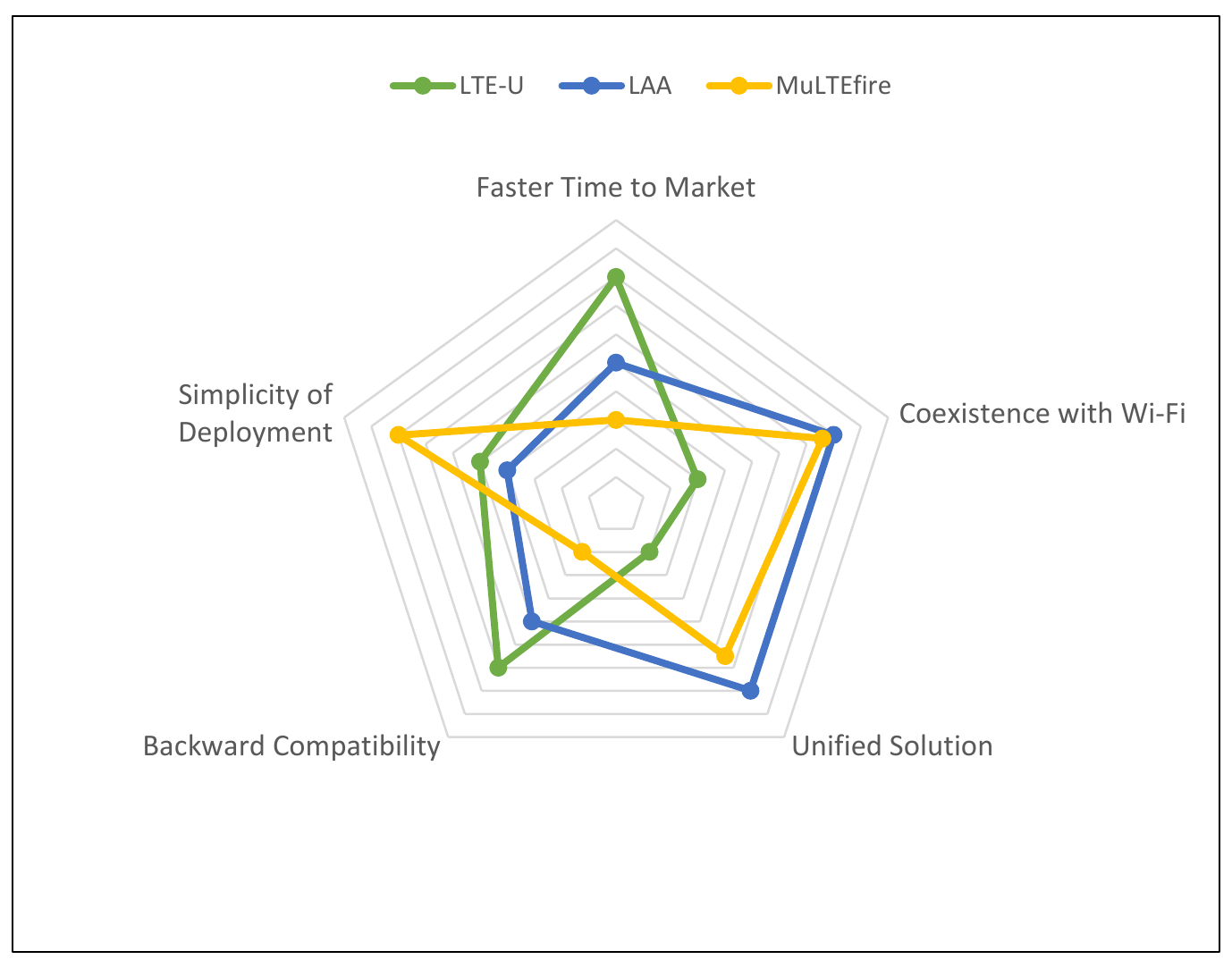} 
		\caption{Comparison between the three variants of LTE in unlicensed bands.}
		\label{fig:compare}
	\end{center}
	\vspace{-0.25in}
\end{figure}

Table \ref{tab:theeversions} and Figure \ref{fig:compare} provide a comparison of the three proposed variants of LTE in unlicensed bands. Note that the radar chart of Figure \ref{fig:compare} reflects the current state of the art. The shapes may change as research, regulation and standardization evolves.

\begin{table}[h] 
	\centering
	\caption{Comparison between the three different versions of LTE in unlicensed bands}
	\begin{tabular} {|m{2.2cm}|m{4.1cm}|m{4.1cm}|m{4.1cm}|}
		\hline
		& \textbf{LTE-U} & \textbf{LAA}   & \textbf{MulteFire} \\
		\hline
		Devloped By & LTE-U Forum & 3GPP  & MulteFire Alliance \\
		\hline
		Mode of Operation & Supplemental downlink (SDL) only & DL is supported in Rel-13, UL is standardized in Rel-14  & DL and UL (using Time Division Multiplexing (TDD)) \\
		\hline
		Frequency Bands & 5150-5250 MHz and 5725-5850 MHZ, whereas the frequency bands 5250-
		5725 MHz has been reserved for future use. &  5150 MHz- 5925 MHz (Channels 46 and 47)  & The 5 GHz, and the 3.5 GHz in the US. The 1.9 \& 2.4 GHz bands are expected to be supported in subsequent releases\\
		\hline
		Bandwidths Allowed & 20 MHz. & 10 MHz and 20 MHz.  & 10 MHz and 20 MHz. \\
		\hline
		Purpose & Use of the existing LTE specifications in countries that do not impose LBT.& Single unified global framework that complies with regulations in the different world regions. & Standalone version that operates without a primary carrier in licensed band. \\
		\hline
		Coexistence Mechanisms & Carrier selection, on-off switching, Carrier-Sensing Adaptive Transmission (CSAT). & Carrier selection and LBT. 	& LBT (similar to LAA). \\ 
		\hline
		Main Advantage & Will likely be the first version in the market; Samsung and Verzion  announced their plans to implement it in 2016. & Will be a single unified global framework for LTE in unlicensed bands worldwide. & Combines benefits of LTE with simplicity of Wi-Fi. \\
		\hline
		Backward compatibility with 3GPP Rel-9  & \multicolumn{2}{l|}{Yes, since using the licensed spectrum as the primary carrier.} & No. \\
		\hline
		
	\end{tabular}%
	\label{tab:theeversions}%
\end{table}%


\section{Conclusion}

This paper has discussed LTE in unlicensed spectrum as the next big milestone in the evolution of LTE. We have presented the different regulatory requirements for the 5 GHz unlicensed bands in different world regions and analyzed the benefits and the challenges for operating LTE in unlicensed spectrum. We have explained  the technical details of the three proposed variants of LTE for unlicensed spectrum and have compared the coexistence mechanism used in these variants. Our numerical analysis has shown that LAA and MulteFire offer better coexistence to Wi-Fi than LTE-U. 

This paper has highlighted the differences and uniquenesses of each variant of LTE operating in unlicensed spectrum. LAA is the most unified solution and will be operable worldwide. LTE-U is less regulated and thus expected to be introduced first. MulteFire is most flexible and will be as simple to deploy as Wi-Fi. Moreover, MulteFire has the potential to play a big role in the future of wireless communications, especially for enterprises and industrial applications, as several major high-tech companies are currently collaborating to improve performance and simplify deployment and configuration. MulteFire also has the potential to enhance the future of the Internet of Things (IoT), as it can enable new IoT deployment scenarios, hence creating new business opportunities. 

It is expected that LTE in unlicensed bands will become more important and the technology be leveraged for LTE moving into additional unlicensed bands. This can be expected with the ongoing spectrum relocation in the US and worldwide. Moreover, history has shown that popular technologies (such as LTE or Wi-Fi) expand.  For example, Wi-Fi has gone through several generations and expanded from the 2.4 GHz band to the 5 GHz, 60 GHz (WiGig) and former TV bands (Super-Wifi).  We will likely see different variants of LTE in unlicensed bands in the market because of different regulations and incentives. This diversity will help gaining experience for next generation wireless networks. Furthermore, LTE and Wi-Fi will both evolve and once radars are relocated, there will be more room for improvement and efficiency to make better use of spectrum opportunities. LTE in unlicensed spectrum is not competing with Wi-Fi; they complement each other, have their pros and cons and may cooperate in the future.

\newpage
\bibliographystyle{IEEEtran}
\bibliography{references}

\end{document}